\documentstyle[12pt]{article}
\newcommand{\be}{\begin{equation}}
\newcommand{\ee}{\end{equation}}
\newcommand{\ba}{\begin{eqnarray}}
\newcommand{\ea}{\end{eqnarray}}
\newcommand{\NL}{\nonumber \\ }

\begin{document}
\thispagestyle{empty}
\setcounter{page}{0}
 
\ \
 
\vspace{2.0cm}
 
\begin{center}
{\LARGE
{\bf A new effective Lagrangian}
 
\vspace{0.5cm}
 
{\bf for nuclear matter} }
\end{center}
\bigskip
 
\begin{center}
 
{\Large
{\bf T.S. Bir\'o and J. Zim\'anyi\footnote{jzimanyi@sunserv.kfki.hu} }
}
 
{\it MTA KFKI  Research Institute for Particle and Nuclear Physics}
 
{\it Budapest 114., P.O.Box 49, H-1525 Hungary}
 
 
\end{center}


\vspace{1.0cm}
 
\begin{center}
{\large {\bf Abstract}}
\end{center}
\medskip

The relativistic mean field  model, the
Zim\'anyi - Moszkowski (ZM) Lagrangian describes nuclear matter and
stable finite nuclei even in the non-relativistic limit.
It fails, however, to predict the correct
non-relativistic spin-orbit (SO) coupling.
In this paper we improve on this matter by an
additional tensor coupling analogous to the anomalous
gyromagnetic ratio. It can be adjusted to
describe the SO-term without changing the mean field
solution of the ZM-Lagrangian for nuclear matter.

\noindent
 
\vspace{0.5cm}
 
\newpage

{\em Introduction}

\vspace{0.5cm}
The advantages of describing nuclear physics relativistically
are to find connections between different parts of
the non-relativistic mean field nuclear potential,
such as the linear energy dependence emerging from
kinematic features of an effective Dirac equation for
single nucleons or a mild mean potential being balanced
between strong attraction and repulsion.

\vspace{0.5cm}
The most known such model, the Walecka model\cite{1a1,1a2} 
was first intended to be an elementary description,
but it was soon discovered that only an effective
Lagrangian approach is fruitful. Renormalization
problems and divergences of higher meson loop
corrections with a strong effective coupling
revealed that by dropping the requirement of an elementary
approach  a higher degree of freedom in the
choice of the effective Lagrangian can be exploited.

\vspace{0.5cm}
The Walecka model describes the non-relativistic
mean field potential, $(V-S)$, and spin-orbit
coupling $\frac{d}{dr} (V+S)$, both derived
from the same effective Dirac equation. 
Unfortunately it fails to describe compressibility
and a finite nucleus, which remains bound in a
quantum molecular dynamics (QMD) approach.
Simulation of dynamics including heavy ions reactions
at relativistic energies, however, desperately needs 
a Lagrangian, which is relativistic and has a stable
nucleus. Such an effective Lagrangian was proposed
by Zim\'anyi and Moszkowski \cite{1b}  recently.
It works with less repulsion, $V$, with an attractive
mean field, $S$ which is bounded from above by the
free nucleon mass, $M$  and describes
stable finite nuclei\cite{1c1}-\cite{1c5}. Up to now only the too weak
SO-potential was a manque of this model\cite{1d}.

\vspace{0.5cm}
Recent investigations did show that this Lagrangian
has many advantageous properties in describing the
bulk properties of the nuclear matter\cite{2a}-\cite{2c}.
The optical potential derived from the ZM Lagrangian also
agrees with experimental findings\cite{3}.
Finite nuclei described within the framework of this model
are used in QMD calculations\cite{4} as well.

\vspace{0.5cm}
In this letter we show that an additional tensor term
influences the spin - orbit part of the non-relativistic
potential without spoiling the benevolent property of
the ZM Lagrangian: the mild repulsive mean field.
This reduced repulsion is necessary to describe
stable finite nuclei in the quantum molecular
dynamical approach to heavy ion collisions.
Throughout  this paper we use Minkowskian metric with 
the signature $(+,-,-,-)$ and the convention $\hbar=c=1$.

\vspace{1.0cm}
{\em Improved effective Lagrangian}

\vspace{0.5cm}
The Zim\'anyi-Moszkowski (ZM) Lagrangian,
\hbox{${\cal L}_{{\rm ZM}},$} is formulated in terms
of nucleon fields $\psi$, a scalar $\sigma$ and
vector $\omega_{\mu}$ mean fields. The corresponding scalar
and vector coupling constants are $g_s$ and $g_v$, respectively.
Now we add a tensor term to this Lagrangian which couples
the antisymmetric spin current density 
\hbox{$\overline{\psi} \sigma_{\mu\nu} \psi$}
to the field strength of the vector field $F^{\mu\nu}$
\ba
{\cal L}_{{\rm ZMB}} &=& - \overline{\psi} M_* \psi
+ \overline{\psi} \left(
(i\partial_{\mu} - g_v\omega_{\mu}) \cdot \gamma^{\mu}
+ \frac{\lambda}{2M_*} g_v \sigma_{\mu\nu} F^{\mu\nu} \right)
\psi \NL
&&- \frac{1}{4}F_{\mu\nu}F^{\mu\nu}
+\frac{1}{2}m^2\omega_{\mu}\omega^{\mu}
+\frac{1}{2} \partial_{\mu}\sigma \cdot \partial^{\mu}\sigma
-U(\sigma).
\ea
with
\be
M_* = \frac{M}{1+g_s\sigma/M}
\ee
being the effective nucleon mass.
$M$ and $m$ are the {\em bare} nucleon and
vector meson masses, respectively, and $U(\sigma)$ includes
the \hbox{$\frac{1}{2}m_{\sigma}^2\sigma^2$} 
mass term and an eventual self interaction.

\vspace{0.5cm}
In order to deal with static mean fields over a
finite spatial extension we assume
that only $\sigma$ and 
\hbox{$\omega_{\mu}=(\omega_0,\vec{0})$} differs from zero
and all time derivative vanishes. In this description
of static nuclei $\omega_0$ and $\sigma$ may have gradients, however.
The field strength $F_{\mu\nu}$ has only electric components
\hbox{$ F_{0k} =  - \nabla_k \omega_0.$}
The corresponding coefficient Dirac-matrix in the tensor term is 
\hbox{$\sigma_{0k} = - i \alpha_k.$}
The equations of motion for the meson fields can be derived from
the improved Lagrangian easily:
\ba
-\Box\sigma \, - \, U^{\prime}(\sigma) \, + \,
\frac{M_*^2}{M^2} g_s \overline{\psi}\psi \, + \,
\frac{\lambda}{2M} \frac{g_s}{M} g_v 
F^{\mu\nu} \overline{\psi} \sigma_{\mu\nu} \psi
& = & 0, \NL
\partial_{\mu} \left( F^{\mu\nu} + 
\frac{\lambda}{M_*} g_v \overline{\psi}\sigma_{\mu\nu}\psi
\right) \, + \, m^2\omega^{\nu} - g_v
\overline{\psi}\gamma^{\nu}\psi & = & 0.
\ea
These equations lead to the same mean field equations
as without the tensor term for all unpolarized media with 
\hbox{$\langle \overline{\psi}\sigma_{\mu\nu}\psi \rangle = 0.$}
Using this the homogeneous static mean field solution
arises as in the case of the Zim\'anyi - Moszkowski Lagrangian
with $U(\sigma)=\frac{1}{2}m_{\sigma}^2\sigma^2$:
\ba
m_{\sigma}^2 \, \sigma & = & \frac{M_*^2}{M^2} g_s
\langle \overline{\psi}\psi \rangle, \NL
m^2 \, \omega_0 & = & g_v
\langle \psi^{\dag}\psi \rangle.
\ea

\vspace{0.5cm}
The Dirac equation derived from the above Lagrangian
on the other hand has a new contribution
\be
\left( \gamma^{\mu} ( i\partial_{\mu} - g_v\omega_{\mu})
+ \frac{\lambda}{2M_*} g_v \sigma_{\mu\nu} F^{\mu\nu}
- M_* \right) \psi = 0.
\ee
In order to investigate the non-relativistic limit
we consider the Hamiltonian, expressing
$\hat{H}\psi = i\partial_t \psi$ from the above
equation. Multiplication with $\gamma_0$ from the
left and neglecting vanishing mean field and time
derivative components gives the relativistic Hamiltonian 
\be
\hat{H} = - i \vec{\alpha} \vec{\nabla} + g_v\omega_0
- \frac{\lambda}{2M_*} i \vec{\gamma} \vec{\nabla} g_v\omega_0
+ M_* \beta,
\ee
where we used the Dirac representation of matrices
\be
\beta = \left( \begin{array}{cc} 1 & 0 \\ 0 & 1 \end{array} \right),
\qquad
\vec{\alpha} =
\left( \begin{array}{cc} 0 & \vec{\sigma}  \\ \vec{\sigma}  & 0 \end{array} \right),
\qquad
\vec{\gamma} =
\left( \begin{array}{cc} 0 & \vec{\sigma}  \\ -\vec{\sigma}  & 0 \end{array} \right).
\ee
For the sake of a more compact notation we introduce
\hbox{$S = g_s\sigma/(1+g_s\sigma/M)$} and
\hbox{$V = g_v\omega_0.$} 
This notation automatically satisfies the relation
$M_* = M - S$.
The relativistic Hamiltonian equals
\be
\hat{H} = \beta M^* + \vec{\alpha} \vec{p}
+ V - i\frac{\lambda}{2M_*} \vec{\gamma} (\vec{\nabla}V).
\ee
where the momentum operator
$\vec{p} = -i\vec{\nabla}$ is introduced. 
The Dirac matrices and spinors in $2\times 2$ notation,
\hbox{$\psi = (\varphi, \chi)$},
lead to coupled equations for the upper ($\varphi$) and lower ($\chi$)
components. The non-relativistic Hamiltonian is
the effective Hamiltonian acting on the upper
component of the Dirac spinor.
We get the following eigenvalue equations
\ba
\left( V + M^*  - E \right) \varphi
\, + \,  \vec{\sigma} \left( \vec{p} - 
i\frac{\lambda}{2M_*}\vec{\nabla}V \right) \chi &=& 0
\NL
\vec{\sigma} \left( \vec{p} + 
i\frac{\lambda}{2M_*}\vec{\nabla}V \right) \varphi 
 \, + \, \left( V - M^*  - E \right) \chi &=& 0,
\ea
with $E$ being the relativistic energy eigenvalue.
The non-relativistic energy eigenvalue is therefore
$e=E-M$. Introducing further simplifying  notations,
\be
\vec{p}_{\pm} = \vec{p} \mp 
i \frac{\lambda}{2M_*} (\vec{\nabla} V), \qquad
V_{\pm} = V \pm M^* - E ,
\ee
it reduces to
\ba
V_+ \varphi \, + \, \vec{\sigma} \vec{p}_+ \chi &=& 0,
\NL
\vec{\sigma}\vec{p}_- \, + \, V_- \chi & = & 0.
\ea
Here $V_{\pm}$ can be expressed by using the non-relativistic
energy eigenvalue as
\hbox{$V_+ = V - S - e$} and
\hbox{$V_- = V - S - e - 2M_*.$}
Extracting now $\chi$ from the second equation and
substituting back into the first gives
\be
\left( V_+ \, - \, 
\vec{\sigma}\vec{p}_+ \frac{1}{V_-} \vec{\sigma}\vec{p}_-
\right) \varphi = 0.
\ee
A non-relativistic Hamiltonian can be obtained 
by expressing the energy eigenvalue $e$ from this equation.
We get
\be
\hat{h}\varphi = e \varphi = 
\left( V - S \, - \,
\vec{\sigma}\vec{p}_+ \frac{1}{V_-} \vec{\sigma}\vec{p}_-
\right) \varphi.
\ee
The non-relativistic expansion of this operator together
with a gradient expansion for the mean field factors
$V$ and $S$ leads to the non-relativistic nuclear
potential. The spin-orbit coupling term arises 
as a sub-leading correction in an ${\cal O}(1/M_*)$
expansion. Traditionally it is due to the fact that
the $\vec{p}$ gradient operator does not commute with
$1/V_-$. In the present case we have an additional
contribution to this term stemming from the tensor
coupling due to 
\hbox{$\vec{p}_+ \times \vec{p}_- \ne 0$.}
We proceed by calculating 
\be
\vec{p}_+ \frac{1}{V_-} =
\frac{1}{V_-}\vec{p}_+ \, - \, i \vec{\nabla}\frac{1}{V_-}.
\ee
After noting that $\vec{\nabla}V_- = \vec{\nabla}(V+S)$ 
both terms on the right hand side can be expanded in inverse
powers of the effective mass. We get
\be
\vec{p}_+ \frac{1}{V_-} =
-\frac{\vec{p}}{2M_*} + 
\frac{1}{4M_*^2} \left( 
\left[ e - (V-S) \right] \vec{p}
+ i\vec{\nabla} \left[ (1+\lambda)V + S \right]
\right) + \ldots
\ee
Let us denote this result by $-\vec{\pi}_+/2M_*$.
Using this notation the non-relativistic Hamiltonian is
\be
\hat{h} = V-S + \frac{1}{2M_*}
\vec{\sigma}\cdot\vec{\pi}_+  \,\,
\vec{\sigma}\cdot\vec{p}_- + \ldots
\label{NRENER}
\ee
with
\be
\vec{\pi}_+ =  \vec{p}
- \frac{1}{2M_*} \left(
\left[ e -(V-S) \right] \vec{p}
+ i \vec{\nabla} \left[ (1+\lambda)V + S \right]
\right) 
\ee

\vspace{0.5cm}
Now we utilize the general algebraic identity,
\be
\vec{\sigma}\cdot\vec{\pi}_+ \,\,\,\, \vec{\sigma}\cdot\vec{p}_- =
\vec{\pi}_+\cdot\vec{p}_- 
+ i \vec{\sigma}\left( \vec{\pi}_+ \times \vec{p}_- \right).
\ee
The individual terms are again expanded
\be
\vec{\pi}_+\cdot\vec{p}_- =
\left( 1 - \frac{e-(V-S)}{2M_*} \right) \vec{p}^{\,\,\, 2}
- \frac{1}{2M_*} \vec{\nabla} (V+S) \cdot \vec{\nabla}
+  \vec{\nabla} \left( \frac{\lambda}{2M_*} \vec{\nabla}V \right)
+ \ldots 
\ee
and
\be
\vec{\pi}_+ \times \vec{p}_- =
\frac{-i}{2M_*} \vec{\nabla} \left[ (1+2\lambda)V + S \right]
\, \times \, \vec{p} \, \,
+ \, \, \vec{\nabla} \frac{\lambda}{2M_*} \, \times \,
\vec{\nabla} V + \ldots 
\ee
The non-relativistic energy eigenvalue of (\ref{NRENER})
in an $1/M_*$ is given by
\be
\hat{h}_0 = V - S + \frac{\vec{p}^{\,\,\, 2}}{2M_*} + \ldots
\ee
Substituting this leading order result into the coefficient
of $\vec{p}^{\,\,\, 2}$ in the next order term we make a
higher order error only. This arises as a relativistic correction to
the kinetic energy
\be
\hat{h}_1 = - \frac{1}{2M_*} \frac{e_0-(V-S)}{2M_*}\vec{p}^{\,\,\, 2}
= - \frac{\vec{p}^{\,\,\, 4}}{8M_*^3} + \ldots
\ee
Two further interesting term arise from the  scalar product
$\vec{\pi}_+\cdot\vec{p}_-$:
\be
\hat{h}_2 = - \frac{1}{4M_*^2} \vec{\nabla} (V+S) \cdot \vec{\nabla}
+ \frac{1}{2M_*} \vec{\nabla} \left( \frac{\lambda}{2M_*} \right)
\vec{\nabla}V .
\ee
Factorizing the $1/4M_*^2$ coefficient we get
\be
\hat{h}_2 = - \frac{1}{4M_*^2} \left(
\vec{\nabla} (V+S) \cdot \vec{\nabla}
\, - \, \lambda M_* \vec{\nabla}
\left( \frac{1}{M_*} \vec{\nabla}V \right)
\right).
\ee
This will be the analogue of the so-called Darwin term.
After symmetrization and integration by part we arrive at
\be
e_2 =  \frac{1}{8M_*^2} \vec{\nabla}^2 \left[
(1+2\lambda)V + S \right]
+ \frac{1}{4M_*^3} (1+\lambda) \vec{\nabla}S \cdot
\vec{\nabla} (V+S)
\ee
The familiar non relativistic Darwin term stems from
a Coulomb potential ($V=V_C$, $S=0$ and $M_*=M$).
In this case $\vec{\nabla}^2 V_C \propto \delta(\vec{r})$
and it can be interpreted as the spin-orbit energy
shift to zero angular momentum states.

\vspace{0.5cm}
Finally the term stemming from the vector product
between $\vec{\pi}_+$ and $\vec{p}_-$ leads to
\be
\hat{h}_3 =
\frac{1}{4M_*^2} \vec{\sigma} \left(
\vec{\nabla}((1+2\lambda)V+S) \, \times \, \vec{p}
\right)
\, + \, i\vec{\sigma} \frac{1}{2M_*}
\left( \vec{\nabla} \frac{\lambda}{2M_*}
\, \times \, \vec{\nabla}V \right).
\ee
Acting by the gradient $\vec{\nabla}$ on the $1/M_*$
coefficient a result higher order in $1/M_*$ arises,
which can be consequently neglected. Finally we arrive at
\be
\hat{h}_3 = \frac{1}{4M_*^2} \vec{\sigma} \left(
\vec{\nabla}((1+2\lambda)V+S) \, \times \, \vec{p}
\right) + \ldots
\ee
This result -- and its derivation -- is analogous to the 
treatment of an anomalous gyromagnetic ratio
\hbox{$(\lambda = \Delta g/4, g_v = e)$},
\cite{IZ}.

\vspace{0.5cm}
For central mean fields $V=V(r)$, $S=S(r)$
the first term of this expression can be rewritten 
by using the angular momentum operator,
$\vec{L}= \vec{r} \times \vec{p}$:
\be
\hat{h}_3 = \frac{1}{4M_*^2}  \left(
\frac{1}{r} \frac{d}{dr} ((1+2\lambda)V+S) \right)
\vec{\sigma}\cdot\vec{L}.
\ee


\vspace{1.0cm}
{\em Conclusion}

\vspace{0.5cm}
Summarizing the main effect of introducing the
direct tensor coupling term in the relativistic
Lagrangian is to enhance the role of the vector
mean field $V$ to $(1+2\lambda) V$ in leading order
spin-orbit  corrections to the non-relativistic
single nucleon energy.  From this result we 
conclude that the effective strength of the
spin-orbit coupling is increased by the new
term in the Lagrangian,
\hbox{$V_{SO} = (1+2\lambda) V + S.$}
This is the main result of this paper.

\vspace{0.5cm}
The equations of motion for the mean fields do include
an additional term due to the tensor coupling. This
has no effect, however, in the mean field approximation
to unpolarized media. Therefore estimating the spin-orbit
potential we can use the values of $V$ and $S$ obtained
self consistently for the ZM Lagrangian ($V=82$ MeV,
$S=140$ MeV).
Comparing the strength of the respective terms with 
that of the Walecka model \hbox{$(V+S=785$ MeV)} 
we obtain $\lambda = 3.5$.
One can improve on this estimate by applying the presently described
Lagrangian, \hbox{${\cal L}_{{\rm ZMB}}$}, for finite nuclei.

\vspace{0.5cm}
It is interesting to note that P.G. Reinhard and his collaborators also
considered a tensorial coupling in connection with the
original Walecka Lagrangian. They discarded, however,
this possibility in that case, because they found its
contribution negligible\cite{6}.

\vspace{1.0cm}
 
{\em Acknowledgment} 

\vspace{0.5cm}
The drilling questions and steady requiring of
Judit N\'emeth and S. Moszkowski, which triggered this work, 
are gratefully acknowledged.
This work was supported by the Hungarian Science Research Fund, OTKA
No.T014213.

\bigskip

\vfill

\end{document}